\documentclass[CRPHYS,Unicode,manuscript]{cedram}

\usepackage{amssymb}
\usepackage{url}
\usepackage{hyperref}

\title{Superfluid transition in quasi-two-dimensional disordered dipolar Fermi gases}

\author{\firstname{Victoria} 
\middlename{Y.} \lastname{Pinchenkova}\IsCorresp}
\address{Russian Quantum Center, Skolkovo, Moscow 143025, Russia}
\address{Moscow Institute of Physics and Technology, Dolgoprudny, Moscow Region, 141701, Russia}
    
\email[V. Y. Pinchenkova]{vpinchenkova@gmail.com}

\author{\firstname{Sergey} \middlename{I.} \lastname{Matveenko}}
\address{L. D. Landau Institute for Theoretical Physics, Chernogolovka, Moscow region 142432, Russia}
\addressSameAs{1}{<repeat address 1>}
\email[S. I. Matveenko]{smatveenko@yahoo.com}

\author{\firstname{Vladimir}
\middlename{I.} \lastname{Yudson}}
\address{Laboratory for Condensed Matter Physics, HSE University, Moscow 101000, Russia}
\addressSameAs{1}{<repeat address 1>}
\email[V. I. Yudson]{v.yudson@googlemail.com}

\author{\firstname{Georgy}
\middlename{V.} \lastname{Shlyapnikov}}
\addressSameAs{1}{<repeat address 1>}
\addressSameAs{2}{<repeat address 2>}
\address{Université Paris-Saclay, CNRS, LPTMS, 91405 Orsay, France}
\address{Van der Waals–Zeeman Institute, Institute of Physics, University of Amsterdam, Science Park 904, 1098 XH Amsterdam, The Netherlands}
\email[G. V. Shlyapnikov]{georgy.shlyapnikov@universite-paris-saclay.fr}


\begin{abstract} 
We investigate the effect of weak disorder on the superfluid properties of two-component quasi-two-dimensional dipolar Fermi gases. The dipole-dipole interaction amplitude is momentum dependent, which violates the Anderson theorem claiming that the weak disorder has practically no influence on the superfluid transition temperature in the weakly interacting regime. We find that for dipolar fermions the transition temperature in this regime can be strongly increased by the disorder like in the purely two-dimensional case. However, the effect becomes smaller with increasing the intercomponent fermion-fermion interaction, and in the strongly interacting regime the superfluid transition temperature in the weak disorder becomes very close to that in the absence of disorder.
\end{abstract}

\begin{document}
\maketitle

\section{Introduction}

The study of ultracold atomic Fermi gases became a rapidly growing domain of theoretical and experimental research \cite{Bloch, Giorgini}. One of the important questions here is the interplay between interactions and disorder (see, e.g., \cite{PL,TS,BGM} and references therein), including the influence of a random potential on the superfluid transition temperature. Anderson demonstrated that in the case of a short-range interparticle interaction a weak disorder practically does not change the Bardeen -- Cooper -- Schrieffer (BCS) transition temperature in a weakly interacting Fermi gas (Anderson theorem) \cite{Anderson}. Abrikosov and Gor’kov explored this problem within the diagrammatic approach \cite{AG}. They showed the applicability of the Anderson theorem in the leading order of the parameter $1/(k_F l) \ll1$, where $k_F$ is the Fermi momentum and $l$ is the mean free path. The effect of the weak disorder on the superfluid transition temperature was also evaluated beyond the mean-field approach of Abrikosov and Gor’kov \cite{dis1,dis2,dis3,dis4,dis5,dis6,dis7, dis8}. In this case the Anderson theorem does not hold, as well as it does not hold in the presence of a strong disorder (see, e.g., \cite{Feigel, Burmistrov}). However, this theorem can be violated even on the mean-field level, for example, by the momentum dependence of the fermion-fermion scattering amplitude. As was pointed out in the recent work \cite{Superfluid}, the weak disorder can have a significant effect on the superfluid transition temperature of two-dimensional (2D) dipolar fermions with weak interactions.

The papers cited above mostly cover weakly interacting Fermi systems. However, the realization of Feshbach resonances encouraged the investigation of Fermi gases in the strongly interacting regime. In this regime the superfluid transition in ultracold Fermi gases has been identified experimentally in three-dimensional (3D) \cite{Bloch, Giorgini} and in 2D \cite{expQ2D1} geometries. Theoretical description of 3D strongly interacting Fermi gases on the mean-field level was proposed by Leggett \cite{Leggett1, Leggett2}. The results of this method are compatible with Monte-Carlo calculations performed in a later stage \cite{Bloch, Giorgini}. An acceptable accuracy of this approach in 3D encourages its application to 2D Fermi gases, which in the absence of disorder has been started by Miyake \cite{Miyake}. In the 2D geometry the mean-field predictions also agree reasonably well with the Monte-Carlo \cite{MC1} and the experimental \cite{expQ2D2} results.




In this paper we investigate the effect of weak disorder on quasi-2D dipolar Fermi gases, moving from the weakly to strongly interacting regime. As usual, the term quasi-2D means that in two directions the motion of particles is free, and in the third direction it is harmonically confined to zero point oscillations. We consider two-component ultracold dipolar fermions with dipoles perpendicular to the plane of their translational motion and confine ourselves to the case where the intercomponent interaction amplitude is negative (fermion-fermion attraction). This can be magnetic atoms or polar molecules, for example, a mixture of fermionic isotopes of dysprosium in the lowest Zeeman states. We explore the Berezinskii -- Kosterlitz -- Thouless (BKT) superfluid transition \cite{Berezinskii, Kosterlitz} in such systems relying on the mean-field Leggett approach \cite{Leggett1, Leggett2} and treating the disorder effects in the framework of the theory of Abrikosov and Gor’kov \cite{AG} \footnote{We omit beyond mean-field weak localization effects because in the weakly interacting regime and on approach to the strongly interacting regime they are small compared to corrections caused by the momentum-dependent dipole-dipole interaction amplitude \cite{Superfluid}}.

We find that although the weak disorder potential can strongly increase the BKT superfluid transition temperature $T_{BKT}$ of a weakly interacting dipolar Fermi gas, with increasing the interaction strength the effect of the disorder on $T_{BKT}$ decreases. In the strongly interacting regime the influence of the momentum-dependent component of the dipole-dipole interaction on $T_{BKT}$ is almost unaffected by the weak disorder.

The paper is organized as follows. Sec. \ref{sec2} is devoted to general relations for superfluid pairing in 2D Fermi systems in the presence of weak disorder. In Sec. \ref{sec3} we discuss the s-wave scattering amplitude in quasi-2D dipolar Fermi gases. The main equations for the BKT superfluid transition are given in Sec. \ref{sec4}, where we present the final results for the disorder-induced change of $T_{BKT}$. Our conclusions are given in Sec. \ref{sec5}.

\section{Superfluid pairing in 2D disordered Fermi gases. General relations}\label{sec2}

We first present relations for a two-component purely 2D Fermi gas in a disorder potential $U(\mathbf{r})$. The Hamiltonian is given by

\begin{equation}\label{eq1}
\begin{gathered}
\hat{H} = \int d^2 \mathbf{r} \sum_{\alpha = \uparrow, \downarrow} \hat{\Psi}^{\dagger}_\alpha(\mathbf{r}) \left ( -\frac{1}{2 m} \nabla^2 - \mu +U(\mathbf{r}) \right )\hat{\Psi}_\alpha(\mathbf{r}) +\int d^2 \mathbf{r} \: d^2 \mathbf{r'} \: \hat{\Psi}^{\dagger}_{\uparrow} (\mathbf{r}) \hat{\Psi}^{\dagger}_{\downarrow} (\mathbf{r'}) V(\mathbf{r -r'})\hat{\Psi}_{\downarrow} (\mathbf{r'}) \hat{\Psi}_{\uparrow} (\mathbf{r}) ,
\end{gathered}
\end{equation}
where $\mathbf{r}$ is the 2D coordinate, $\hat{\Psi}_{\uparrow} (\mathbf{r})$ and $\hat{\Psi}_{\downarrow} (\mathbf{r})$ are the field operators of fermionic components, $m$ is the particle mass, $\mu$ is the chemical potential, and we put $\hbar = 1$. The pairing is due to an effective attractive interaction between the fermions characterized by the interaction potential $V(\mathbf{r -r'})$. 

For weakly interacting systems it is common to use the mean-field BCS theory to deal with Hamiltonian (\ref{eq1}). Also, it was shown that one can use the BCS-like approach to qualitatively recover the physics in the strongly interacting regime in 3D \cite{Leggett1,Leggett2} and in 2D \cite{Miyake}. We thus will use the mean-field theory for both weakly and strongly interacting regimes. In the framework of this theory, the superfluid phase is characterized by the order parameter (gap) $\Delta(\mathbf{r}, \mathbf{r'})$:

\begin{equation}\label{gap}
\Delta(\mathbf{r}, \mathbf{r'}) = V(\mathbf{r -r'}) \left< \hat{\Psi}_{\downarrow} (\mathbf{r'}) \hat{\Psi}_{\uparrow} (\mathbf{r}) \right>,
\end{equation}
where the symbol $\left< ...\right>$ denotes the statistical average. 

The superfluid properties are conveniently described by the Green functions formalism. In the absence of disorder ($U(\mathbf{r}) = 0$) we rewrite the Hamiltonian (\ref{eq1}) in terms of annihilation and creation operators $\hat{a}_{\mathbf{k} \alpha}$ and $\hat{a}^{\dagger}_{\mathbf{k}\alpha}$ of fermions with 2D momentum $\mathbf{k}$. We then turn to the Heisenberg representation with operators $\hat{a}_{\mathbf{k}\alpha} (\tau) = e^{\hat{H} \tau} \hat{a}_{\mathbf{k}\alpha} e^{-\hat{H} \tau}$ and $\hat{a}^{\dagger}_{\mathbf{k}\alpha} (\tau) = e^{\hat{H} \tau} \hat{a}^{\dagger}_{\mathbf{k}\alpha} e^{-\hat{H} \tau}$, where $\tau$ is an imaginary time. In this representation the normal and anomalous finite-temperature Green functions are given by $G(\mathbf{k}, \tau) =  -\left<T_\tau \hat{a}_{\mathbf{k}\alpha} (\tau) \: \hat{a}^{\dagger}_{\mathbf{k}\alpha} (0) \right>$ and $F(\mathbf{k}, \tau) =  -\left<T_\tau \hat{a}_{\mathbf{k}\uparrow} (\tau) \: \hat{a}_{-\mathbf{k}\downarrow} (0) \right>$, respectively, where $T_\tau$ is the time-ordering operator. In uniform gases the order parameter (\ref{gap}) depends on the coordinates only through the difference $(\mathbf{r -r'})$. 
After straightforward calculations we find that in the frequency and momentum space the anomalous
Green function is related to the Fourier transform of the superfluid order parameter $\Delta_\mathbf{k}$ by the gap equation

\begin{equation}\label{eq2}
\Delta_\mathbf{k} =  -T \sum_{\omega_j} \int \frac{d^2 \mathbf{k'}}{(2 \pi)^2} V(\mathbf{k'} - \mathbf{k}) F (\mathbf{k'}, \omega_j),
\end{equation}
where $\omega_j = \pi T(2j+1)$, $j = 0, \pm 1, ...$ are the fermion Matsubara frequencies.

In the presence of weak disorder (such that $k_F l \gg 1$) we will modify the main equations, including the gap equation (\ref{eq2}), within the mean-field Abrikosov and Gor’kov method \cite{AG}. According to this approach, one should use Green functions $\left < G (\mathbf{k}, \omega_j)\right>_{dis} \equiv \bar{G}(\mathbf{k}, \omega_j)$ and $\left < F (\mathbf{k}, \omega_j) \right>_{dis} \equiv \bar{F}(\mathbf{k}, \omega_j)$  averaged over the disorder instead of usual ones and take into consideration only self-energies due to the disorder scattering $\Sigma_G(\mathbf{k}, \omega_j)$ and $\Sigma_{F}(\mathbf{k}, \omega_j)$ in averaged Green functions. We consider the short-range disorder potential $U(\mathbf{r})$ with the correlation function $\left < U(\mathbf{r}) U(\mathbf{r'})\right >_{dis} = \gamma \delta (\mathbf{r}-\mathbf{r'})$. With such short-range potential, in the frequency and momentum space the self-energies $\Sigma_G$ and $\Sigma_F$ are momentum independent and given by

\begin{equation}\label{eq3}
\Sigma_G(\omega_j) = \gamma \int \frac{d^2 \mathbf{k}}{(2 \pi)^2} \bar{G}(\mathbf{k}, \omega_j),
\end{equation}

\begin{equation}\label{eq4}
\Sigma_{F}(\omega_j) =   \gamma \int \frac{d^2 \mathbf{k}}{(2 \pi)^2} \bar{F}^{\dagger} (\mathbf{k}, \omega_j).
\end{equation}
Taking into account the disorder contribution, we arrive at Gor'kov equations \cite{Gorkov} in the form

\begin{equation}\label{eq5}
(i \omega_j - \xi_k - \Sigma_G) \bar{G} + (\Delta_\mathbf{k} + \Sigma_{F}) \bar{F}^{\dagger}  =  1,
\end{equation}

\begin{equation}\label{eq6}
(i \omega_j + \xi_k - \Sigma_G) \bar{F}^{\dagger} + (\Delta_\mathbf{k} + \Sigma_{F}) \bar{G} =  0,
\end{equation}
where we omitted arguments ($\mathbf{k}$, $\omega_j$) for brevity, assumed that $\Delta_{\mathbf{k}} = \Delta_{\mathbf{k}}^*$, and put $\xi_k = k^2/(2m) - \mu$. The solutions of equations (\ref{eq5}) and (\ref{eq6}) are

\begin{equation}\label{eq7}
\bar{G}(\mathbf{k}, \omega_j) = - \frac{i \omega_j - \Sigma_G + \xi_k}{ -(i \omega_j - \Sigma_G)^2 + \xi_k^2 + (\Delta_\mathbf{k} + \Sigma_{F})^2},
\end{equation}

\begin{equation}\label{eq8}
\bar{F}^{\dagger} (\mathbf{k}, \omega_j) = \frac{\Delta_\mathbf{k} + \Sigma_{F}}{ -(i \omega_j - \Sigma_G)^2 + \xi_k^2 + (\Delta_\mathbf{k} + \Sigma_{F})^2}.
\end{equation}
Expressing the order parameter $\Delta_\mathbf{k}$ in terms of the states with orbital quantum numbers $\tilde{m}$, we have $\Delta_{\mathbf{k}} = \sum_{-\infty}^{\infty} \Delta_{\tilde{m}} (k) \: e^{i \tilde{m} \phi_k}$. At ultralow energies we confine ourselves to the s-wave scattering. Hence, $\Delta_{\mathbf{k}} = \Delta(k)$ (where we denote $\Delta(k) \equiv \Delta_{\tilde{m} =0} (k)$), and the Green functions do not depend on the polar angle.

We then substitute Eqs. (\ref{eq7}) and (\ref{eq8}) into the relations for self-energies (\ref{eq3}) and (\ref{eq4}). The main contribution to the integrals comes from $k$ close to $k_\mu = \sqrt{2 m \mu}$ and, hence, we can put $\Delta_\mathbf{k} = \Delta(k_\mu)$ under the integrals. We notice that there is a constant term arising in the self-energy $\Sigma_G$. It can be considered as a renormalization of the chemical potential, and it is given by $\delta \mu = \frac{1}{2 \pi \tau_e} \textup{ln}(\Lambda/\mu)$, where we put a finite upper bound $\Lambda \sim \mu$ in the integral over $\xi_k$ because this integral is logarithmically divergent. The quantity $\tau_e$ is the time between disorder-induced elastic collisions in the Born approximation, and $1/\tau_e = 2 \pi \nu \gamma$ with $\nu = m/(2\pi)$ being the 2D density of states.

Subtracting the constant term $\delta \mu$ from the self-energy $\Sigma_G$, we thus find 

\begin{equation}\label{eq9}
\Sigma_G(\omega_j) = -\frac{i}{2\tau_e} \frac{\omega_j}{ \sqrt{ \omega_j^2 + \Delta^2}},
\end{equation}

\begin{equation}\label{eq10}
\Sigma_{F} (\omega_j) = \frac{1}{2\tau_e} \frac{\Delta}{ \sqrt{ \omega_j^2 + \Delta^2}},
\end{equation}
where $\Delta \equiv \Delta(k_\mu)$. Substituting the self-energies into Eqs. (\ref{eq7}) and (\ref{eq8}) we obtain relations for the Green functions averaged over the disorder:

\begin{equation}\label{eq11}
\bar{G}(\mathbf{k},\omega_j) = - \frac{i  \tilde{\omega}_j + \xi_k}{ \tilde{\omega}^2_j + \xi^2_k + \tilde{\Delta}^2_k },
\end{equation}

\begin{equation}\label{eq12}
\bar{F}(\mathbf{k},\omega_j) =  \frac{\tilde{\Delta}_k}{ \tilde{\omega}^2_j + \xi^2_k + \tilde{\Delta}^2_k },
\end{equation}
where $\tilde{\omega}_j$ and $\tilde{\Delta}_k$ are related to the fermionic Matsubara frequency $\omega_j$ and the gap $\Delta(k)$ as

\begin{equation}\label{eq13}
\tilde{\omega}_j = \omega_j \left ( 1 + \frac{1}{2 \tau_e \sqrt{\omega_j^2 + \Delta^2} } \right ),
\end{equation}

\begin{equation}\label{eq14}
\tilde{\Delta}_k= \Delta(k) + \frac{\Delta}{2 \tau_e \sqrt{\omega_j^2 + \Delta^2} }.
\end{equation}
Thus, according to the Abrikosov-Gor'kov approach in the presence of weak disorder the gap equation is given by Eq. (\ref{eq2}), but with $F(\mathbf{k},\omega_j)$ replaced by $\bar{F}(\mathbf{k},\omega_j)$ (\ref{eq12}). 

We now rewrite the gap equation averaged over the disorder in the form

\begin{equation}\label{eq15}
\Delta_\mathbf{k} =  -\int \frac{d^2 \mathbf{k'}}{(2 \pi)^2} V(\mathbf{k'} - \mathbf{k}) \Delta_{\mathbf{k'}} K(k'),
\end{equation}
where the function $K(k')$ comes from the anomalous Green function $\bar{F}(\mathbf{k},\omega_j)$ (\ref{eq12}) and is given by

\begin{equation}\label{eq16}
K(k') = T \sum_{\omega_j}\frac{\tilde{\Delta}_{k'}}{\Delta(k')(\tilde{\omega}^2_j +  \xi^2_{k'}+\tilde{\Delta}^2_{k'})}.
\end{equation}
Eq. (\ref{eq15}) should be renormalized to circumvent the divergence. The renormalized gap equation can be found with the help of the relation between the fermion-fermion scattering amplitude $f(\mathbf{k'}, \mathbf{k})$ and the Fourier transform of the interaction potential $V(\mathbf{k'- k})$ \cite{Landau}:

\begin{equation}\label{eq17}
f(\mathbf{k', k}) = V(\mathbf{k'- k}) + \int \frac{d^2 \mathbf{q}}{(2 \pi)^2} \frac{V(\mathbf{k'- q}) \: f(\mathbf{q, k})}{2(E_k - E_q - i0)},
\end{equation}
with $E_k = k^2/(2m)$. We then multiply this relation by $K(k') \Delta_{\mathbf{k'}}$ and integrate over $d^2 \mathbf{k'}$. Taking into account Eq. (\ref{eq15}), we find

\begin{equation}\label{eq18}
\Delta(k) = - \textup{P} \int \frac{k' d k' }{2 \pi} f(k' ,k) \Delta(k') \left [ K(k') - \frac{1}{2 (E_{k'} - E_k)}\right ],
\end{equation}
where $f(\mathbf{k'}, \mathbf{k})$ is replaced by its s-wave part $f(k',k)$, and the symbol P denotes the principal value. 

The main conclusions of this section, such as the form of Green functions (\ref{eq11}-\ref{eq14}) and the renormalized gap equation averaged over the disorder (\ref{eq18}), are also valid for quasi-2D Fermi gases, but with the quasi-2D attractive scattering amplitude $f(\mathbf{k'}, \mathbf{k})$ discussed in the next section.

\section{Scattering in quasi-2D dipolar Fermi gases}\label{sec3}

We now consider a quasi-2D dipolar Fermi gas tightly confined in the $z$ direction by the harmonic potential $V_H(z) = m \omega_0^2 z^2/4$. The gas is called tightly confined if the characteristic transverse size $l_0 = \sqrt{1/(m \omega_0)}$, is much smaller than the mean interparticle separation $n^{-1/2}$, where $n$ is the particle density. This requirement results in the condition $k_F l_0 \ll 1$ with the Fermi momentum $k_F = \sqrt{2 \pi n}$.

In the case of a quasi-2D Fermi gas with a short-range contact and long-range dipole-dipole interactions an effective 2D s-wave scattering amplitude consists of two terms:

\begin{equation}\label{eq19}
f(k',k) = F_0 + f_{dd} (k', k).
\end{equation}
The local term of the scattering amplitude $F_0$ can be put momentum independent \cite{Superfluid} and can be varied by an external magnetic field due to Feshbach resonances. The nonlocal (dipole-dipole) term $f_{dd} (k', k)$ depends on the momentum, which violates the Anderson theorem. The dipole-dipole part of the scattering amplitude in the Born approximation is given by

\begin{equation}\label{eq20}
f_{dd} (\mathbf{k}', \mathbf{k})= \int d^2 \mathbf{r} \: dz  \: \left \{ \textup{exp}[i \mathbf{r} (\mathbf{k}' - \mathbf{k}) ] - 1 \right \}  V_{dd} (\mathbf{r}, z) \phi^2(z).
\end{equation}
The second term in the curly brackets comes from the fact that the local part of the dipole-dipole scattering amplitude is already included in $F_0$. When the particle dipoles are perpendicular to the plane of their translational motion, and the dipole moments $d_0$ of different components are equal to each other, the dipole-dipole interaction potential $V_{dd} (\mathbf{r}, z)$ reads

\begin{equation}\label{eq21}
V_{dd} (\mathbf{r}, z) = d_0^2 \left ( \frac{ 1}{(r^2 + z^2)^{3/2}} - \frac{3 z^2}{(r^2 + z^2)^{5/2}}\right ). 
\end{equation}
The ground state wavefunction $\phi(z)$ for the harmonic confining potential $V_H(z)$ is 

\begin{equation}\label{eq22}
\phi (z) = \frac{1}{ (2 \pi l_0^2)^{1/4}} \textup{exp} \left ( -\frac{z^2}{4 l_0^2} \right).
\end{equation}
Integrating Eq. (\ref{eq20}) over $z$ and keeping only the s-wave part we find

\begin{equation}\label{eq23}
f_{dd} (k', k) =  \int_{0}^{\infty} \frac{d_0^2}{r^3} \left ( J_0(k r) J_0 (k' r) - 1 \right ) \Phi(r) 2 \pi r dr, 
\end{equation}
with $J_0$ being the Bessel function. The function $\Phi(r)$ is given by

\begin{equation}\label{eq24}
\Phi (r) =  \frac{r}{2 \sqrt{2} l_0} \left [2 U(\frac{1}{2}, 0, \frac{r^2 }{2 l_0^2}) - 3 U(\frac{3}{2}, 0, \frac{r^2}{2 l_0^2}) \right], 
\end{equation}
where $U$ is the Tricomi confluent hypergeometric function. In the limit $l_0 \rightarrow 0$ we have $\Phi(r) = 1$ and, hence, Eq. (\ref{eq23}) recovers the known expression for the purely 2D case \cite{Superfluid}. The validity of Eq. (\ref{eq23}) requires the condition $k_F r_* \ll1$ with $r_* = m d_0^2$ being the so-called dipole-dipole distance. 

\section{BKT transition in the weakly and strongly interacting regimes}\label{sec4}

We assume that the interaction is weak if the parameter $\lambda = |f(k_F, k_F)|m/(2 \pi) \ll 1$, and the interaction is strong if $\lambda$ approaches unity, where the attractive scattering amplitude $f  (k_F,k_F) < 0$ is given by Eqs. (\ref{eq19}) and (\ref{eq23}).

In the theory of superfluidity in ultracold quantum Fermi gases the main processes usually take place in the vicinity of the Fermi surface. Thus, for further calculations we will use an approximate ansatz for the order parameter $\Delta(k') = \Delta(k_\mu)$. For $k = k_\mu$ we get the gap equation (\ref{eq18}) in the form

\begin{equation}\label{eq25}
1 =  - \textup{P} \int \frac{k 'd k'}{2 \pi} f  (k',k_\mu) \left [K(k') - \frac{1}{2 (E_{k'} - \mu)} \right ].
\end{equation}
This equation allows us to find the critical BCS transition temperature $T_{c}$ determined by the condition $\Delta = 0$. However, in 2D as well as in quasi-2D the temperature for the onset of superfluidity differs from the BCS $T_c$, and we should consider the BKT superfluid transition at $T_{BKT} < T_c$. At $T_{BKT}$ the superfluid density $n_s$ jumps from zero to the value given by the Nelson-Kosterlitz relation \cite{Nelson}

\begin{equation}\label{eq26}
n_s = \frac{8 m}{\pi} T_{BKT}.
\end{equation}
In the clean case the superfluid density at a given temperature reads (see, e.g., Ref \cite{Coleman})

\begin{equation}\label{eq27}
n_s(T) = \frac{2}{m} T \sum_{\omega_j} \int \frac{d^2 \mathbf{k}}{(2 \pi)^2} k^2 |F(\mathbf{k}, \omega_j)|^2.
\end{equation}
In the presence of weak disorder, according to the Abrikosov-Gor’kov theory, we should modify the expression for $n_s$ by replacing the anomalous
Green function $F(\mathbf{k},\omega_j)$ with the averaged one $\bar{F}(\mathbf{k},\omega_j)$ (\ref{eq12}):

\begin{equation}\label{eq28}
n_s(T) = \frac{2}{m} T \sum_{\omega_j} \int \frac{ k d k}{2 \pi}   \frac{k^2\tilde{\Delta}_k^2}{ (\tilde{\omega}^2_j + \xi^2_k + \tilde{\Delta}^2_k)^2}.
\end{equation}
Complementing equations (\ref{eq25}), (\ref{eq26}) and (\ref{eq28}) with the condition $\mu = \varepsilon_F$ (with $\varepsilon_F$ being the Fermi energy) and the anzatz $\Delta(k) = \Delta(k_F)$ we obtain self-consistent equations for $\Delta$ and $T_{BKT}$ in the weakly interacting regime. We have solved them numerically (see Fig. \ref{fig1}, \ref{fig2} and discussion below). 

\begin{figure}[tbp]
\includegraphics[width=0.85\linewidth]{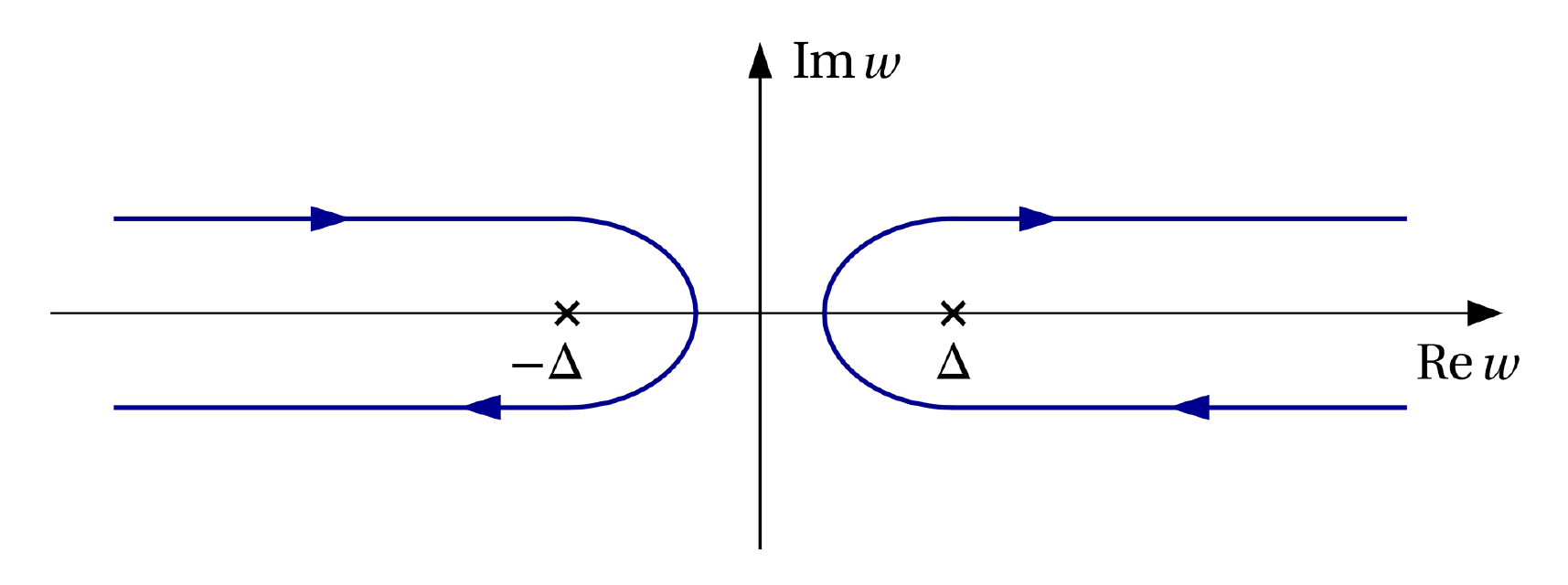}
 \caption{The deformed integration contour in the complex plane $w$ employed in calculating the sum in Eq. (\ref{eq29}).}
 \label{contour}
 \end{figure}

\begin{figure}[tbp]
 \includegraphics[width=0.95\linewidth]{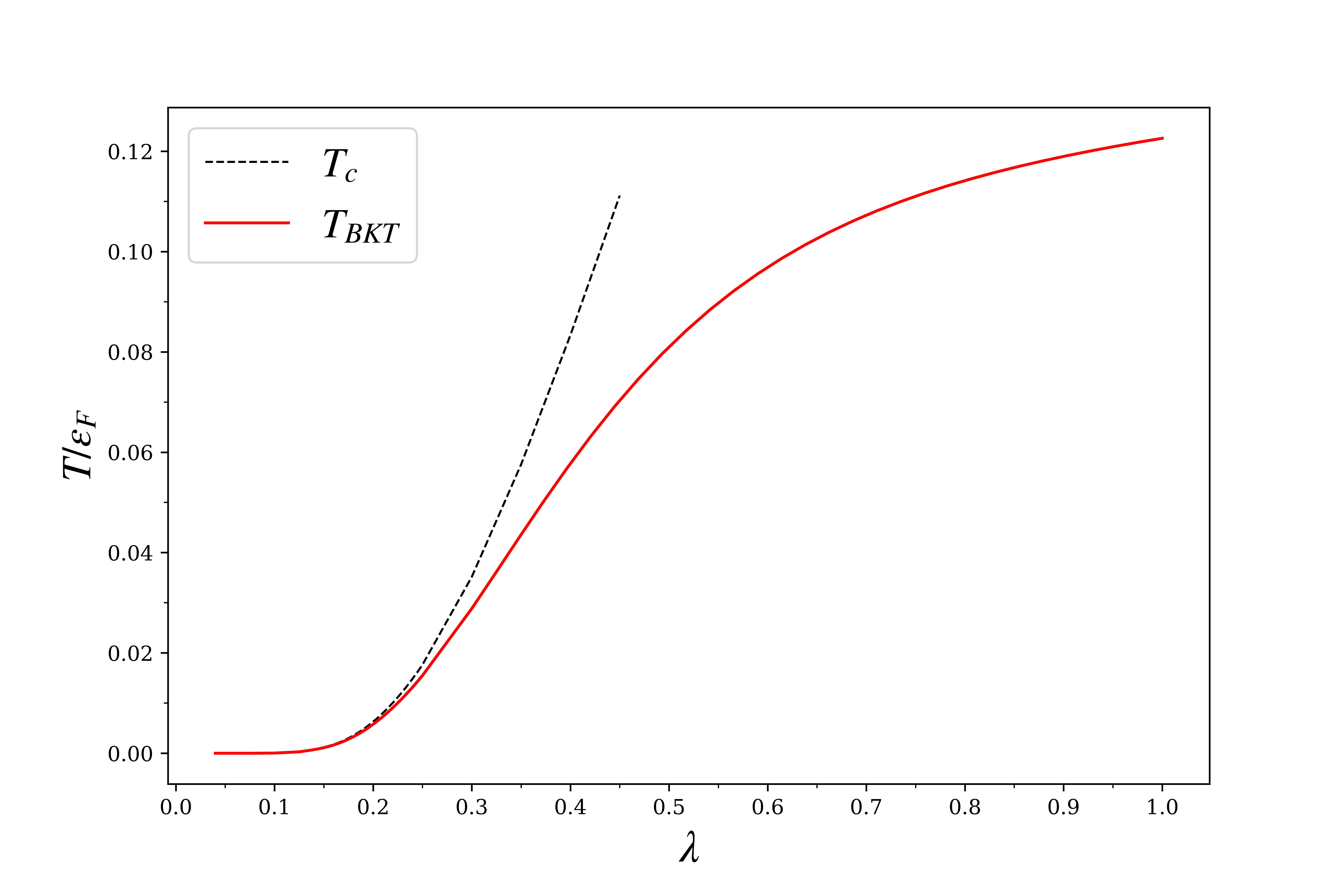}
 \caption{The critical BCS temperature $T_c$ (black dashed line) and the BKT temperature $T_{BKT}$ (red solid line) in the presence of weak disorder versus $\lambda$ for $k_F l = 20, \: k_F r_* = 0.1$ and $k_F l_0 = 0.01$.}
 \label{fig1}
 \end{figure}

\begin{figure}[tbp] \includegraphics[width=0.95\linewidth]{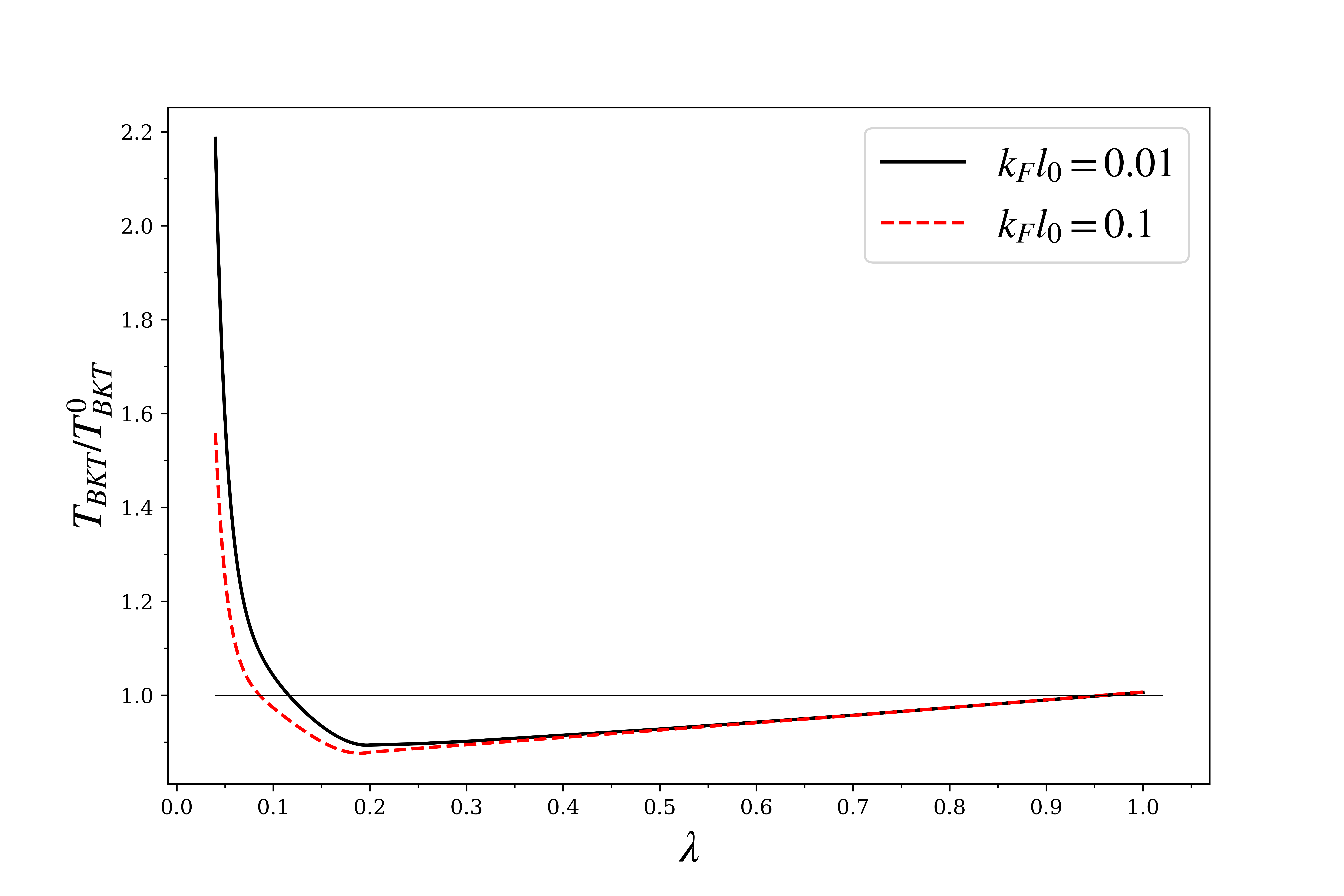}
 \caption{The disorder-induced change of the BKT temperature $T_{BKT}$ versus $\lambda$ with $T_{BKT}^0$ being the BKT temperature in the absence of disorder. $k_F l$ = 20 and $k_F r_*$ = 0.1.}
 \label{fig2}
 \end{figure}

In the intermediate and strongly interacting regimes we can not assume that the chemical potential $\mu$ is equal to the Fermi energy. Equations (\ref{eq25}), (\ref{eq26}) and (\ref{eq28}) for $\Delta$ and $T_{BKT}$ should be complemented with the normalization condition to determine the value of $\mu$. In the frequency-momentum representation and in the presence of disorder the normalization condition is given by

\begin{equation}\label{eq29}
n = 2 T \sum_{\omega_j} \int \frac{d^2 \mathbf{k}}{(2 \pi)^2} \bar{G} (\mathbf{k}, \omega_j) e^{i \omega_j \varepsilon} \left.  \right |_{\varepsilon \rightarrow + 0}.
\end{equation}
Taking into account the relation for the averaged Green function $\bar{G} (\mathbf{k}, \omega_j)$ (\ref{eq11}) with the ansatz $\Delta(k) = \Delta(k_{\mu})$ we, as usual, change the frequency sum in Eq. (\ref{eq29}) to a contour integral in the complex plane $w$: 

\begin{equation}\label{eq30}
\left. \int \: \frac{d w}{2 \pi i} \frac{w( 1 + \frac{1}{2 \tau_e g(w)})+ \xi_k}{[ g(w) + 1/(2 \tau_e) ]^2 +  \xi_k^2}  n_F(w) e^{w \varepsilon} \right |_{\varepsilon \rightarrow + 0},
\end{equation} 
with $n_F(w) = (\textup{exp}(w/T) + 1)^{-1}$. The branch of the function $g(w) = \sqrt{\Delta^2 - w^2}$ is determined by the branch cuts $w \in (-\infty, -\Delta)$, $w \in (\Delta, \infty)$ and the condition $g(0) = \Delta$. Deforming the integration contour as shown in Fig. \ref{contour} in order to exclude the branch cuts we find

\begin{equation}\label{eq31}
\begin{gathered}
n = \frac{m}{2 \pi} \int_{-\mu}^{\Lambda} d\xi_k  \int_{\Delta}^{\infty} \frac{d x}{ \pi} \left [ B_{-} (x, k) + \textup{Tanh}\frac{x}{2T} B_{+} (x, k) \right ],
\end{gathered}
\end{equation}
where $x = $Re$(w)$, $B_{\pm}(x, k) = \textup{Im} (B(-x, k) \pm B(x, k))$ and

\begin{equation}\label{eq32}
B(x, k) =  \frac{ x(1 + \frac{1}{2\tau_e i \sqrt{x^2-\Delta^2}}) + \xi_k }{\left (i \sqrt{x^2-\Delta^2}  + 1/(2 \tau_e) \right )^2 
 + \xi_k^2  }.
\end{equation}

Strictly speaking, in the presence of disorder the integral over $\xi_{k}$ in Eq. (\ref{eq31}) is logarithmically divergent. To deal with this problem we should put a finite upper bound $\Lambda \sim \varepsilon_F $ in the integral over $\xi_k$. We verified numerically that the value of $\Lambda$ practically has no influence on $T_{BKT}$ in the intermediate and strongly interacting regimes. The difference between the values of $T_{BKT}$ with $\Lambda = 2 \varepsilon_F$ and with $\Lambda = 5 \varepsilon_F$ is less than 5 percent for all considered values of $k_F l_0$.

Fig. \ref{fig1} compares the exact transition temperature $T_{BKT}$ with the critical BCS temperature $T_c$ moving from the weakly to strongly interacting regime for $k_F l = 20$, $k_F r_* = 0.1$ and $k_F l_0 = 0.01$. We see that in the weakly interacting regime the BKT temperature $T_{BKT}$ is very close to the BCS $T_c$. Numerical calculations have shown that for $\lambda \lesssim 0.15$ the difference between these two temperatures is less than 1 percent in the absence of disorder (such that $k_F l \rightarrow \infty$) and less than 6 percent in the presence of weak disorder (such that $k_F l = 20$) for all considered values of $k_F l_0$. Hence, in this limit we can put $T_{BKT} \approx T_c$, which is in agreement with Ref. \cite{Miyake}. Moving away from the limit $\lambda \ll1$ the approximate BCS temperature can no longer be interpreted as the transition temperature from the normal to superfluid state, and we should consider the exact value of $T_{BKT}$.

As shown in Fig. \ref{fig1}, for $\lambda= 0.3$ the ratio $T_{BKT}/\varepsilon_F$ can reach 0.03. For typical 2D densities $n \sim 10^9$ cm$^{-2}$ the Fermi energy of dysprosium isotopes is of the order of 100 nK. Thus, the superfluid transition temperature for $\lambda= 0.3$ can reach $10$ nK and higher, which is realistic for current experiments with ultracold Fermi gases \cite{Turlapov}. With increasing $\lambda$ the BKT temperature monotonically increases, and, hence, can also be achieved in experiments.

\newpage

Fig. \ref{fig2} shows the ratio of the BKT transition temperature in the presence of disorder to that temperature in the clean case $T_{BKT}/T_{BKT}^0$ versus $\lambda$ for $k_F l$ = 20, $k_F r_*$ = 0.1 and for various values of $k_F l_0$. In the limit $\lambda \ll 1$ and in the case of the strong confinement (such that $k_F l_0 = 0.01$) the interplay between the dipole-dipole interaction and the weak disorder leads to a significant increase of $T_{BKT}$ like in the purely 2D case \cite{Superfluid}. For weaker confinement the effect of the disorder becomes smaller but is still significant. With an increase in the interaction strength the influence of weak disorder on $T_{BKT}$ decreases and practically vanishes in the strongly interacting regime even in the case of the strong confinement.

\section{Concluding remarks}\label{sec5}
In conclusion, we have analyzed the effect of weak disorder on the BKT superfluid transition temperature of weakly and strongly interacting quasi-2D dipolar Fermi gases. We have obtained within the mean-field theory that the disorder-induced corrections to $T_{BKT}$ strongly depend on the interaction strength. In the weakly interacting regime the superfluid transition temperature in weak disorder can significantly increase compared to $T_{BKT}^0$ in the clean case. In the strongly interacting regime $T_{BKT}$ in the weak disorder is very close to $T_{BKT}^0$ without disorder. Also, in the weakly interacting regime the disorder-induced corrections depend on the confinement, namely, the influence of the disorder increases with increasing the confinement.

Our results for the intermediate and strongly interacting regimes can be tested in experiments, for example, with a two-component gas of dysprosium isotopes, $^{161}$Dy and $^{163}$Dy, in the lowest Zeeman sublevels and with equal concentrations of the components. Their magnetic moments can be oriented perpendicularly to the plane of the translational motion, and one can neglect a small difference in masses of the isotopes. The superfluid transition temperature of such system can be several nanokelvins and higher for realistic parameters. Temperatures close to these values have already been achieved in experiments with ultracold Fermi gases \cite{Turlapov}. For example, for the density of dysprosium isotopes $n = 6 \times 10^8$ cm$^{-2}$ we find the Fermi momentum $k_F \approx 6 \times 10^4$ cm$^{-1}$ and the Fermi energy $\varepsilon_F = \pi n/m \approx 60$ nK. The magnetic moment of the dysprosium isotopes is equal to 10$\mu_B$, and the dipole-dipole distance is $r_* \approx 200$ angstroms, so that $k_F r_* \approx 0.1$. The strong confinement can be realized by a trapping frequency $\omega_0 = 2 \pi \times 100$ kHz, for which $k_F l_0 \approx 0.1$. Selecting the mean free path $l = 3 \times 10^{-4}$ cm we obtain $k_F l \approx 20$. For these parameters in the strongly interacting regime the BKT temperature is $T_{BKT} \approx 0.12 \varepsilon_F \approx 7$ nK. Increasing the density to $n = 2 \times 10^9$ cm$^{-2}$ we find $T_{BKT} \approx 23$ nK.

\section*{Acknowledgments}
We are grateful to Murod Bahovadinov for useful comments regarding numerical calculations. This work was supported by the Russian Science
Foundation Grant No. 20-42-05002. V. Y. also acknowledges Basic research program of HSE.

\bibliographystyle{crunsrt}

\nocite{*}

\bibliography{samplebib}

\end{document}